\documentclass[prb,twocolumn,longbibliography,amsmath,superscriptaddress,amssymb,nobibnotes]{revtex4-2}

\usepackage{graphicx}
\usepackage{dcolumn}
\usepackage{bm}
\usepackage{amssymb}
\usepackage{amsmath}
\usepackage{color}
\usepackage[dvipsnames]{xcolor}
\usepackage{placeins}
\usepackage{epstopdf}
\usepackage{multirow}
\usepackage{fontenc}
\usepackage{tabularx}
\usepackage[normalem]{ulem}
\usepackage[linkcolor=blue,urlcolor=blue,citecolor=blue,colorlinks=true]{hyperref}

\newcommand{\ecp}{{EuCd$_2$P$_2$}}
\newcommand{\ezp}{{EuZn$_2$P$_2$}}

\begin{document}

\title{Magnetism, heat capacity and electronic structure of EuCd$_2$P$_2$ in view of its colossal magnetoresistance}

\author{Dmitry~Yu.~Usachov}
\email{dmitry.usachov@spbu.ru}
\affiliation{St. Petersburg State University, 7/9 Universitetskaya nab., St. Petersburg, 199034, Russia}
\affiliation{Moscow Institute of Physics and Technology, Institute Lane 9, Dolgoprudny, Russia}
\affiliation{National University of Science and Technology MISIS, Moscow, 119049 Russia}

\author{Sarah~Krebber}
\affiliation{Physikalisches Institut, Goethe-Universit\"at Frankfurt, 60438 Frankfurt/M, Germany}

\author{Kirill~A.~Bokai}
\affiliation{St. Petersburg State University, 7/9 Universitetskaya nab., St. Petersburg, 199034, Russia}
\affiliation{Moscow Institute of Physics and Technology, Institute Lane 9, Dolgoprudny, Russia}

\author{Artem~V.~Tarasov}
\affiliation{St. Petersburg State University, 7/9 Universitetskaya nab., St. Petersburg, 199034, Russia}
\affiliation{Moscow Institute of Physics and Technology, Institute Lane 9, Dolgoprudny, Russia}

\author{Marvin~Kopp}
\affiliation{Physikalisches Institut, Goethe-Universit\"at Frankfurt, 60438 Frankfurt/M, Germany}

\author{Charu~Garg}
\affiliation{Physikalisches Institut, Goethe-Universit\"at Frankfurt, 60438 Frankfurt/M, Germany}

\author{Alexander~Virovets}
\affiliation{Institute of Inorganic Chemistry, Goethe-Universit\"at Frankfurt, 60438 Frankfurt/M, Germany}

\author{Jens~M\"uller}
\affiliation{Physikalisches Institut, Goethe-Universit\"at Frankfurt, 60438 Frankfurt/M, Germany}

\author{Max~Mende}
\affiliation{Institut f\"ur Festk\"orper- und Materialphysik, Technische Universit\"at Dresden, Dresden D-01062, Germany}

\author{Georg~Poelchen}
\affiliation{Institut f\"ur Festk\"orper- und Materialphysik, Technische Universit\"at Dresden, Dresden D-01062, Germany}

\author{Denis~V.~Vyalikh}
\affiliation{Donostia International Physics Center (DIPC), 20018 Donostia-San Sebastian, Spain}
\affiliation{IKERBASQUE, Basque Foundation for Science, 48013 Bilbao, Spain}

\author{Cornelius~Krellner}
\affiliation{Physikalisches Institut, Goethe-Universit\"at Frankfurt, 60438 Frankfurt/M, Germany}

\author{Kristin~Kliemt}
\email{kliemt@physik.uni-frankfurt.de}
\affiliation{Physikalisches Institut, Goethe-Universit\"at Frankfurt, 60438 Frankfurt/M, Germany}

\date{\today}
\begin{abstract}
The mechanism of the peculiar transport properties around the magnetic ordering temperature  of semiconducting antiferromagnetic EuCd$_2$P$_2$ is not yet understood. With a huge peak in the resistivity observed above the N\'eel temperature, $T_{\rm N}=10.6\,\rm K$, it exhibits a colossal magnetoresistance effect. Recent reports on observations of ferromagnetic contributions above $T_{\rm N}$ as well as metallic behavior below this temperature have motivated us to perform a comprehensive characterization of this material, including its resistivity, heat capacity, magnetic properties and electronic structure. Our transport measurements revealed quite different temperature dependence of resistivity with the maximum at 14\,K instead of previously reported 18~K. Low-field susceptibility data support the presence of static ferromagnetism above $T_{\rm N}$ and show a complex behavior of the material at small applied magnetic fields. Namely, signatures of reorientation of magnetic domains are observed up to $T=16$~K. Our magnetization measurements indicate a magnetocrystalline anisotropy which also leads to a preferred alignment of the magnetic clusters above $T_{\rm N}$. The momentum-resolved photoemission experiments at temperatures from 24\,K down to 2.5\,K indicate the permanent presence of a fundamental band gap without change of the electronic structure when going through $T_N$ that is in contradiction with previous results. We performed \textit{ab initio} band structure calculations which are in good agreement with the measured photoemission data when assuming an antiferromagnetic ground state. Calculations for the ferromagnetic phase show a much smaller bandgap, indicating the importance of possible ferromagnetic contributions for the explanation of the colossal magnetoresistance effect in the related  EuZn$_2$P$_2$.
\end{abstract}

\maketitle
\section{Introduction}

In recent years, materials showing unusual transport properties like anomalous Hall effect or a colossal magnetoresistance (CMR) were studied intensively \cite{PhysRevLett.76.1356, Ramirez_1997, ramirez1997colossal} as they potentially form the materials basis for future applications \cite{osti_1429310, Tokura2017}. It turned out that promising candidates can be found among Eu-based 122 systems with the general formula Eu$T_2X_2$ ($T$ = Cd, Zn and $X$ = P, As, Sb) in the trigonal CaAl$_2$Si$_2$ structure type (space group $P\bar{3}m1$, No.164). It was shown that many of these materials such as EuCd$_2$Sb$_2$ \cite{Soh2018}, EuCd$_2$As$_2$ \cite{Rahn2018}, EuCd$_2$P$_2$ \cite{Sunko2023} and EuZn$_2$P$_2$ \cite{Krebber2023} posses an A-type antiferromagnetic (AFM) structure with in-plane moments. A large anomalous Hall effect and topological properties in its paramagnetic phase were found in EuCd$_2$As$_2$ \cite{Ma2019} and like the As compound, also EuCd$_2$Sb$_2$ is a magnetic Weyl semimetal \cite{Su_EuCd2Sb2_2020}.
EuCd$_2$P$_2$ \cite{Wang2021} and EuZn$_2$P$_2$ \cite{Krebber2023} show a colossal magnetoresistance effect whose origin is still under investigation. 
Previously, such substantial CMR effects were primarily observed in mixed-valent perovskite manganites, rare earth chalcogenides, or europium hexaboride \cite{ramirez1997colossal,TOKURA19991,DAGOTTO20011,Tokura_2006, PhysRevB.94.224404, Krebber2023} among other examples. However, the majority of these systems exhibit ferromagnetic (FM) ordering. With these new Eu-based materials, there is now a growing interest in studying CMR in antiferromagnetically ordered materials.

In EuCd$_2$P$_2$, a pronounced peak in the resistivity appears above $T_N$ which can be suppressed in magnetic field by several orders of magnitude \cite{Wang2021}.
While first results on EuCd$_2$P$_2$ \cite{Wang2021} proposed spin fluctuations being responsible for the unusually high negative MR value in this compound, in a recent study by means of optical polarimetry, static ferromagnetic order above its N\'eel temperature was detected \cite{Sunko2023}.  
This is in contrast to the presence of short-range fluctuations reported in the related material EuCd$_2$As$_2$ \cite{Soh2020}. 
As a possible mechanism that might cause the CMR in EuCd$_2$P$_2$, the formation and percolation of FM clusters was proposed \cite{Sunko2023} which is supported by the results of recent optical conductivity measurements \cite{Homes2023}. 
Sunko et al. \cite{Sunko2023} further propose, that the FM clusters after their formation at temperatures higher than the temperature of the resistivity peak start to merge which results in the decrease of the resistivity upon further cooling. 
From these studies, the following picture can be drawn: In the layered antiferromagnet EuCd$_2$P$_2$, the maximum in the resistivity and the concomitant carrier localization is caused by FM domains which start forming below $\approx$~2$T_{\rm N}$ and result in spin-polarized clusters due to the coupling of localized Eu$^{2+}$ spins and itinerant carriers.
Recent electrical transport data of Eu$_5$In$_2$Sb$_6$ \cite{Rosa_npjQM_2020} show that this Eu-based material is insulating and exhibits an exceptionally large negative magnetoresistance, which is consistent with the presence of magnetic polarons. In {\ecp} a similar CMR effect is observed which strengthens the argument that coupling of Eu$^{2+}$ spins to itinerant carriers is fundamental to the CMR response in Eu-based semiconductors.
In \cite{Sunko2023}, a lower bound of the FM volume fraction was estimated to be 0.1\% which needs to be critically considered in view of the possibility of a temperature-induced spacial overlap of FM clusters. We therefore decided to investigate the magnetic properties of the material in more detail to shed further light on the formation and percolation of the putative FM clusters in EuCd$_2$P$_2$ and provide additional insights if and how the CMR effect might be connected to the reorientation of the magnetic moments in the material upon exposing it to an increasing magnetic field. In related Eu-based antiferromagnetic trigonal and tetragonal systems with in-plane moments, the reorientation of magnetic moments was studied recently in \cite{Pakhira2022, Pakhira2023} and it was found, that in these cases fields of the order of $0.01-0.1\,\rm T$ are needed to induce a realignment of the moments in the $a-a$ plane. In EuCd$_2$P$_2$, the peak in the resistivity can be suppressed already by such low fields and the question arises, whether a reorientation of antiferromagnetic domains occurs in the same field range and might also influence the formation of ferromagnetic clusters.

Here, we present a detailed study of the magnetic properties of {\ecp} in connection with its heat capacity, resistivity and CMR effect. Also in understanding of transport properties an essential point is how the electronic structure of material changes in the temperature range where the CMR effect is observed. Therefore, we perform a photoemission study of the electronic band structure accompanied by density functional theory (DFT) calculations.

\section{Methods}

\subsection{Crystal growth} 

\begin{figure*}[htbp]
\centering
\includegraphics[width=1.0\textwidth]{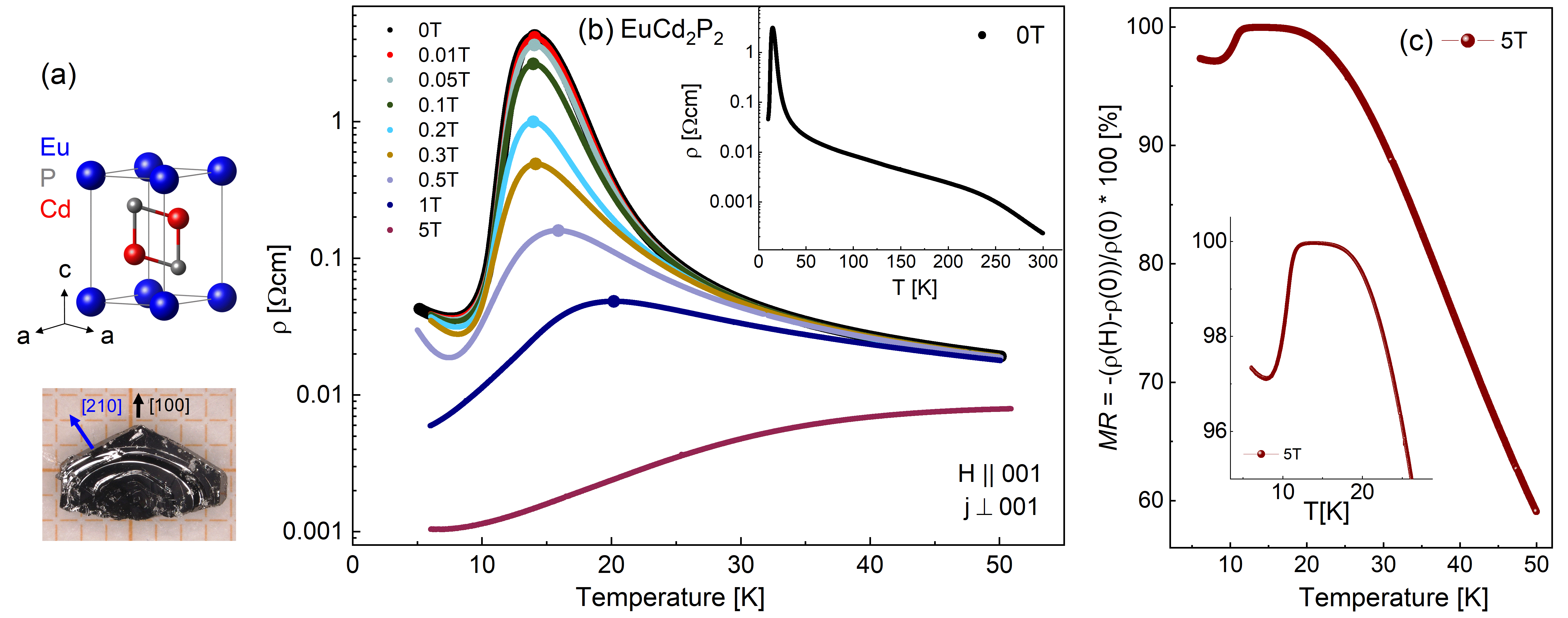}
\caption{\label{fig:EuCd2P2_rho_Mueller}(a) Unit cell of the trigonal crystal structure and as-grown single crystal. (b) Temperature-dependent resistivity measured in the $a-a$ plane with an applied magnetic field oriented along the $c$ axis. The resistivity is strongly suppressed under the application of magnetic field. Large symbols mark the peak position (maximum) of the resistivity. Inset shows the zero-field resistivity up to room temperature. (c) Magneto-resistance (MR), calculated using the formula MR= $[\rho(H)-\rho(0)]/\rho(0)$, evaluated for a magnetic field of $\mu_0H=5$\,T. Notably, a substantial negative MR of 99.96$\%$ is observed at 14\,K, followed by a gradual decline below 8\,K, see inset for an enhanced view of the low-temperature behavior. Subsequently upon cooling, there is a resurgence in the MR signal, also evident in the resistivity measurements.
} 
\end{figure*}

Single crystals of EuCd$_2$P$_2$, Fig.~\ref{fig:EuCd2P2_rho_Mueller}(a), were grown from an external Sn flux by using ingots of europium (99.99\,\%, Evochem), red phosphorous (99.9999\,\%, Chempur), tin (99.999\,\%, Evochem) and teardrops of cadmium (99.9999\,\%, Chempur). The elements were cut into pieces and mixed together with a stoichiometry of Eu:Cd:P:Sn = 1:2:2:20 under an inert Ar atmosphere inside a glove box. The materials were then put in a graphite crucible inside an evacuated quartz ampule. We found that alumina crucibles tend to oxidize the materials easier and therefore used graphite crucibles instead. The ampule was then loaded into a box furnace (Thermconcept), heated up to $450\,^\circ\text{C}$ and held for $5\,\rm h$. This ensures that the phosphorous slowly reacts with the other elements. Afterwards, the temperature was raised to $850\,^\circ\text{C}$ and held for a few hours in order to homogenize the melt. The temperature was then slowly lowered to $600\,^\circ\text{C}$ with a rate of $2\,\text{K}$/h, where the Sn flux was removed by centrifuging. Hexagonal-shaped single crystals with an average size of $3\,\rm mm\times 3\,\rm mm \times 1\,\rm mm$ were obtained. The trigonal crystal structure (space group: No. 164) was confirmed using powder X-ray diffraction (XRD), single crystal XRD, Laue diffraction and energy-dispersive X-ray spectroscopy (EDX). The characterization of the crystal structure of our EuCd$_2$P$_2$ single crystals by powder XRD yields lattice parameters of $a= 4.324$\,\AA\, and $c= 7.179$\,\AA\, which are comparable with the published data \cite{Wang2021}. The powder diffraction patterns were recorded on a diffractometer with Bragg-Brentano geometry and copper K$_\alpha$ radiation (Bruker D8). The orientation of the single crystals was determined by the Laue method using an X-ray equipment ``Micro 91'' (M{\"u}ller) with a tungsten anode. The edges of the crystal correspond to the [210]-direction in the direct lattice while the [100]-direction corresponds to the corner of the crystal [see Fig.\ref{fig:EuCd2P2_rho_Mueller}(a)]. The chemical composition Eu:Cd:P=$(17\pm 3)$:$(40\pm 2)$:$(43\pm 3)$ of the EuCd$_2$P$_2$ single crystals was determined using energy dispersive x-ray spectroscopy (EDX) and is comparable with the expected 1:2:2 stoichiometry of the compound taking a systematic error of our equipment of 2-3\,at\% into account. We also investigated the variation of the stoichiometry of crystals from different batches and found small statistical errors of $(17.0 \pm 0.4)\,\text{at}\%$ for Eu, $(39.6 \pm 0.2)\,\text{at}\%$ for Cd and $(43.5 \pm 0.6)\,\text{at}\%$ for P indicating a high homogeneity of the samples. To exclude that the large absolute error bars in the EDX analysis result from  misoccupation of the different crystallographic sites and to exclude the incorporation of Sn into the crystal structure, we additionally performed a single crystal analysis.

\subsection{Single crystal analysis}
Single-crystal diffraction data were collected at 213\,K on \textit{STOE IPDS II} two-circle diffractometer equipped with the Genix 3D HS microfocus Mo \textit{K$_\alpha$} X-ray source ($\lambda$ = 0.71073 \AA). The finalization of the data, including empirical absorption corrections, was done using the CrysAlisPro software (Rigaku Oxford Diffraction, 2022). The initial structural model was taken from those for EuZn$_2$P$_2$ \cite{Krebber2023}. The structure was refined in the anisotropic approximation against $|F|^2$ with full-matrix least-squares techniques using the program SHELXL-2018/3 \cite{Sheldrick2015}. CIF files containing the crystallographic information are deposited in the Cambridge Crystallographic Data Centre under the deposition code CSD78987. Crystallographic data and parameters of the diffraction experiments are given in Tab.~\ref{tab:SingleCrystalDiff}.

    \vspace*{-\baselineskip}
    \begin{center}
    \begin{table}[!ht]
    \caption{Crystallographic data and refinement results for EuCd$_2$P$_2$ from single crystal X-ray diffraction.}
    \vspace{0.2cm}
    \label{SingleCrystalDifftable}
    \begin{tabularx}{\linewidth}{ll} 
    \hline
    Chemical formula & EuCd$_2$P$_2$\\
    \hline
    \hline
    M$_r$ & 438.70 \\ 
    Crystal system  &Trigonal \\ 
    Space group & P$\overline{3}$m1\,(\text{No.}\,164) \\
    Temperature (K) & 213(1) \\
    $a$ (\,\AA\,) & 4.32421(11) \\
    $c$ (\,\AA\,) & 7.1683(1) \\
    $V$ ({\,\AA\,}$^3$) & 116.08(1) \\
    Z & 1 \\
    \textit{F}(000) & 189 \\
    \textit{D$_x$} (Mg m$^{-3}$) & 6.276 \\
    Radiation type & Mo \textit{K$\alpha$} \\
    $\mu$ (mm$^{-1}$) & 22.90 \\
    Crystal size (mm) & $0.06 \times 0.06 \times 0.04$ \\
    Crystal shape, colour & Triangular plate, black \\
    \textit{R$_{\text{int}}$} & 0.053 \\
    2$\Theta_{max}$ ($^\circ$)	 & 65 \\
    \textit{$R(F)^a, wR(F^2)^b, GooF^c$ \qquad \qquad} & 0.014, 0.035, 1.21 \\
    No. of parameters&  10 \\
    Extinction coefficient & 0.197(8) \\
    $\Delta \rho_{max}$,  $\Delta \rho_{min}$ ($ e \text{\,\AA\,}^{-3}$)& 1.25, -1.57 \\
    \hline
    \multicolumn{2}{l}{}\\
    \multicolumn{2}{l}{$^a R(F) = \sum||F_o|-|F_c||/\sum|F_o|$ for \textit{F$^2$}$ >2\sigma (F^2)$} \\
    \multicolumn{2}{l}{$^bwR(F^2) = [\sum w({F_o}^2-{F_c}^2)^2/\sum w({F_o}^2)]^{1/2}$} \\
    \multicolumn{2}{l}{$^cGooF = [(\sum w({F_o}2{F_c}^2)^2)/(N_{\text{ref}}-N_{\text{param}})]^{1/2}$} \\
    \multicolumn{2}{l}{for all reflections}\\
    \end{tabularx}
    \label{tab:SingleCrystalDiff}
    \end{table}
    \end{center}
    \vspace*{-\baselineskip}

\subsection{Electrical transport}
The electrical transport measurements on the EuCd$_{2}$P$_{2}$ samples were carried out using a standard four-point configuration. Low-resistance electrical contacts were made by thermal evaporation and deposition of 200~nm gold with a 7~nm chromium wetting layer onto the polished surface of the crystals. The electric connection then was made using gold wire and silver paste. In order to verify ohmic behaviour, $I$-$V$ characterization was performed using low-frequency lock-in technique at various temperatures. The resistivity was calculated using the contact geometry and dimensions of the $a-a$ plane and the $c$ axis.

\subsection{Heat capacity and magnetization}
The heat capacity (HC) measurements were carried out using the standard semiadiabatic relaxation method implemented in the Physical property measurement system (PPMS) by Quantum Design. The samples were placed on a calibrated sample puck with Apiezon N grease. At each temperature set point three decay curves were measured applying a heat pulse of 2\,\%. HC curves were measured at different magnetic fields up to 2\,T. The sample was first cooled in zero-field to 2\,K before turning on the magnetic field. The data were then collected while heating the sample to 30\,K, where the magnetic field was removed. Magnetic susceptibility, $\chi(T)$, at fixed magnetic fields over the range of $T=2-300$\,K and magnetic moment versus field, \textit{M(H)}, at constant temperatures were measured using the vibrating sample magnetometry option of the PPMS by Quantum Design. For low fields, $\chi(T)$ was measured in zero-field cooled (ZFC) and field-cooled (FC) mode. Before each ZFC measurement, the temperature was set to $T=35\,\rm K$. Afterwards the temperature was lowered to 2\,K before applying a respective magnetic field. The data were then collected while heating the sample. The FC curves were subsequently obtained by cooling the sample down with the applied magnetic field. We needed to correct the sample mass, due to tin inclusions in the measured crystals. The amount of tin was determined by analyzing the drop in $\chi(T)$ at $T_{SC}= 3.7$\,K. The amount was between 2 and 10 wt.\% and the utilized sample mass was corrected accordingly. At very small fields, the signal is strongly influenced by the background contributions stemming from the sample holder and the glue (GE-varnish) and we therefore show the low-field data in arbitrary units.

\subsection{DFT$+U$ calculations}
\textit{Ab initio} electronic structure calculations were performed using the OpenMX code (ver.~3.9), which implements a fully relativistic norm-conserving pseudopotentials~\cite{Troullier_PRB_1993} and basis functions as linear combination of localized pseudoatomic orbitals~\cite{Ozaki_PRB_2003,Ozaki_PRB_2004}. Calculations were performed using local density approximation (LSDA-CA in OpenMX) for the exchange-correlation energy~\cite{LDA_CA_1980,LDA_PZ_1981}. The basis functions were set as follows: P$7.0-s^2p^2d^1f^1$, Cd$7.0-s^3p^2d^2$, Eu$8.0-s^3p^2d^2f^1$ (the pseudopotential cutoff radius in a.u. is followed by a basis-set specification). The strongly localized Eu $4f$ electrons were treated using the effective Hubbard potential $U_{eff}=2.4$~eV within a simplified approach proposed by Dudarev \textit{et al.}~\cite{Dudarev_PRB_1998}. The cutoff energy of 200~Ry was used to control the accuracy of the real-space numerical integration and solution of the Poisson equation. The total energy convergence criterion was set to $10^{-7}$~eV. Experimental lattice parameters were used to construct the unit cell. Atomic positions in the unit cell were relaxed until the forces on each atom were less than $0.0005$~Hartree/Bohr ($\approx 10^{-2}$~eV/\AA).

\subsection{Angle-resolved photoemission spectroscopy}
ARPES experiments were performed at the U125-2-PGM beamline of BESSY~II synchrotron at the One-Cube end-station. Atomically clean
surface of the \ecp\ crystals were achieved by \textit{in situ} cleavage at around 40~K while the pressure in the vacuum chamber was better than $10^{-10}$~mbar. 

\begin{figure*}[htbp]
\centering
\includegraphics[width=1.0\linewidth]{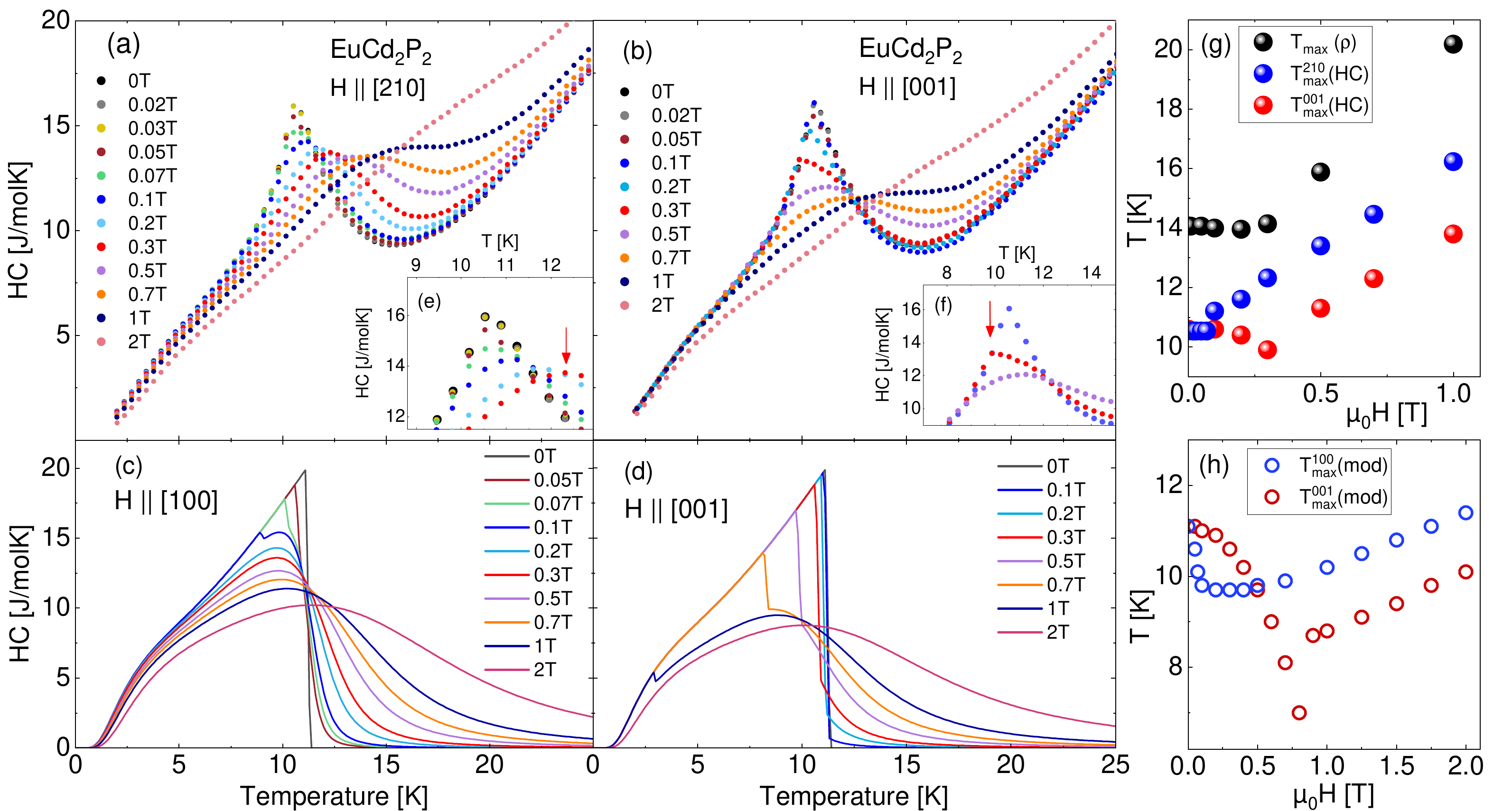}
\caption{\label{fig:EuCd2P2_HC_comp} EuCd$_2$P$_2$ (a) Heat capacity measured with $H\parallel [210]$ 
and (b) with $H\parallel [001]$. 
The modeling of the heat capacity data is shown in (c) for $H\parallel[100]$ and in (d) for $H\parallel[001]$.
(e) No shift of the maximum is observed for $H\leq 0.03\,\rm T$, $H\parallel[210]$. The arrow in the inset (f) shows that a kink appears for $\mu_0H=0.3\,\rm T$, $H\parallel[001]$. The position of the maxima in $\rho$, HC (g) and in the calculated data (h) is shown in for comparison.}
\end{figure*}

\section{Results and discussion}

\subsection{Electric transport measurements}
The temperature-dependent resistivity is shown in Fig.~\ref{fig:EuCd2P2_rho_Mueller}(b) in a semi-logarithmic plot for various magnetic fields applied along the $c$ axis. Starting from its room temperature value of $0.23\,\rm m\Omega\,\rm cm$, see inset, the resistivity of EuCd$_{2}$P$_{2}$ shows a monotonic increase of several orders of magnitude with decreasing temperature until a pronounced maximum reaching a value of about 4\,$\Omega\,$cm occurs at $T_{\rm max} = 14\,\rm K$.
Upon further cooling, there is a pronounced drop in resistance preceding the antiferromagnetic ordering temperature of $T_{N} = 10.6$\,K. This trend continues until about 7\,K, below which the resistance starts to increase again with decreasing temperature. A previous study by Wang {\it et al.} \cite{Wang2021} reported a similar low-temperature resistance profile for this compound, albeit with a slightly broader peak shape and a higher temperature of the resistivity maximum of { $T_{\rm max} = 18$\,K} while the ordering occurs at {$T_{N} = 11.3$\,K} \cite{Wang2021}. However, whereas a metallic behavior of $\rho(T)$. i.e.\ a positive temperature coefficient ${\rm d}\rho/{\rm d}T > 0$, is reported in \cite{Wang2021,Sunko2023} for temperatures down to about 80 - 120\,K, depending on the sample, we find a negative temperature coefficient, ${\rm d}\rho/{\rm d}T < 0$ from room temperature down to $T_{\rm max}$, consistent with the band structure calculations and ARPES measurements discussed below. These sample-to-sample dependencies are currently not understood and require a more detailed investigation of the relationship between growth conditions and electric transport properties that is beyond the scope of this study. 
We note that the low-temperature behaviour about the antiferromagnetic transition observed for EuCd$_{2}$P$_{2}$ is notably distinct from the related compound from the same family of Eu-based 122 systems, EuZn$_{2}$P$_{2}$, where no peak in the resistivity occurs but rather a monotonic increase of the resistivity upon cooling through $T_N$ down to the lowest measured temperatures \cite{Krebber2023}.

In EuCd$_{2}$P$_{2}$ the application of magnetic field along the $c$ direction results in a substantial reduction in resistance, as illustrated in Fig.~\ref{fig:EuCd2P2_rho_Mueller}(b), which for $\mu_0 H = 5$\,T sets in at temperatures as high as 200\,K (not shown). The magnetoresistance (MR), defined as $[\rho(H)-\rho(0)]/\rho(0)$, is negative and takes on rather large values at low temperatures. For instance, at a temperature of 30\,K (which is about twice $T_{\rm max}$) at small fields $\approx 0.2$\,T, the MR amounts to $-32\,\%$ and reaches a value of $-93\,\%$ at 1\,T. These changes become more notable as the temperature is further reduced. The temperature of the largest negative MR coincides with the temperature of the resistivity maximum at $T_{\rm max}$ where the MR increases from $-12.5\,\%$ at $\mu_0 H = 0.05$\,T to $-99.3\,\%$ at 1\,T. Such a behaviour has been reported previously \cite{Wang2021} and is reminiscent of the CMR observed in the related EuZn$_{2}$P$_{2}$ compound. The resistivity peak at around 14\,K in zero magnetic field shifts to higher temperatures, broadens, and nearly levels out when a magnetic field of 5\,T is applied, with a maximum CMR of $-99.96\,\%$ at 5\,T as shown in Fig.~\ref{fig:EuCd2P2_rho_Mueller}(c). Below this temperature range, the negative MR shows a gradual decline until reaching 8\,K below which it begins to increase once more, extending to the lowest temperature of our measurement, a trend that is reflected in the resistivity curves exhibiting an upturn at low temperatures.

\subsection{Single crystal analysis}
The samples investigated in \cite{Wang2021} show a resistivity peak at $18\,\rm K$ while for the samples studied here we find it at $14\,\rm K$. 
This finding might result from a non-stoichiometry of the sample and/or some tiny symmetry distortion, either point (e.g. lowering the crystallographic class) or translational (e.g. modulation). To double-check this, we performed single-crystal XRD. The diffraction pattern can be fully indexed in the hexagonal unit cell without any traces of modulation. There is also no evidence for lowering of either translational or point symmetry. In general, the structure is consistent with one determined by powder XRD \cite{Wang2021} as well as with the structure of Zn-based analog, EuZn$_2$P$_2$ \cite{Krebber2023}, however, leaving some space for possible non-stoichiometry. Unfortunately, refinement of the site occupancy factors (SOFs) is significantly impeded by high X-ray absorption and significant extinction (see Tab.~\ref{tab:SingleCrystalDiff}). At that, the refinement of SOFs for all three atoms (Eu, Cd, P)  based on the diffraction intensities results in the values equal to 100\% within the confidence intervals. To conclude, we can exclude strong structural disorder, the crystalline compound is stoichiometric, and its crystal structure is the one previously reported for EuCd$_2$P$_2$.

\subsection{Heat capacity}
The heat capacity measured with $H\parallel [210]$, Figs.~\ref{fig:EuCd2P2_HC_comp}(a) and S1, shows a pronounced anomaly with a peak at $T=10.6\,\rm K$ which exhibits no apparent shift upon variation of the field for $\mu_0H\leq 0.1\,\rm T$ (inset in Fig.~\ref{fig:EuCd2P2_HC_comp}(a)). 
The comparison with the reported REXS results \cite{Sunko2023}(supplement) shows that the peak in the HC in zero field agrees well with the temperature of the onset of AFM order. Thus we identify this temperature, $T_{\rm N}=10.6\,\rm K$, with the Néel temperature of EuCd$_2$P$_2$.
Upon increasing field, a continuous shift of the maximum to higher temperatures follows. In contrast, when applying the field $H\parallel [001]$, Fig.~\ref{fig:EuCd2P2_HC_comp}(b), it becomes obvious that the heat capacity peak is a superposition of two different contributions. For this field direction, we observe almost no change of the peak position for $\mu_0H\leq 0.1\,\rm T$, but when increasing the field up to $0.3\,\rm T$ the shape of the peak changes as shown (red arrow) in the inset (f) of Fig.~\ref{fig:EuCd2P2_HC_comp} and the peak maximum shifts below 10~K. A similar feature is not observed for $H\parallel [210]$ as it can be seen in the inset (e, red arrow). 

To distinguish unconventional features in the heat capacity and magnetic properties of {\ecp}, we perform a comparison of experimental data with a simple mean-field model that is described further in the text. Figures~\ref{fig:EuCd2P2_HC_comp}(c) and \ref{fig:EuCd2P2_HC_comp}(d) show the model heat capacity for two directions. Note that the model has no in-plane anisotropy and describes only the properties of the $4f$ states. In general, the model heat capacity contains two contributions related to the splitting of the $4f$ states by the external field and by the exchange magnetic field. The exchange field vanishes at $T_{\rm N}$ producing a sharp kink. With the increase of the external field this kink shifts to lower temperatures and disappears since the external field suppresses the AFM order induced by the exchange field. While the kink shifts to lower temperatures quickly for low in-plane fields, the suppression needs slightly higher fields for $H\parallel[001]$. Upon further increase of the field, a broad peak develops due to the field-induced splitting of the $4f$ states. This peak gets broader and shifts to higher temperatures upon the increase of the splitting. Such behavior qualitatively correlates with the experimental data for the [001] direction. However, for the [210] direction the shift of the sharp kink is not resolved; it smoothly transforms into a broad peak as it would be expected for a ferromagnetic material.
 
The field dependence of the heat capacity maximum is shown in Fig.~\ref{fig:EuCd2P2_HC_comp}(g). The experimental data for the two different field directions indicate that the material behaves isotropically for $\mu_0H<0.05\,\rm T$, followed by an increase of the anisotropy between $0.05\leq\mu_0H<0.3\,\rm T$. Above $\mu_0H=0.3\,\rm T$, we find an anisotropy of the peak positions between the two field directions of $\Delta T\approx 2\,\rm K$. It is worth noting that the shift of the heat capacity maximum is much more significant in experiment than in the model, Fig.~\ref{fig:EuCd2P2_HC_comp}(h). Partially, this can be due to the sloped background of other than $4f$ contributions to the heat capacity. A more essential reason can be related to some additional exchange magnetic contributions above $T_{\rm N}$.

In Fig.~\ref{fig:EuCd2P2_HC_comp}(g) we compare the peak position in the heat capacity with that of resistivity.  The peak in the resistivity (Fig.~\ref{fig:EuCd2P2_rho_Mueller}(b)) appears at $14\,\rm K$ for fields $\mu_0H\leq 0.2\,\rm T$. For higher fields, the maximum follows the same trend as the maximum in the heat capacity for $\mu_0H\parallel[001]$. Above $\mu_0H=0.3\,\rm T$, it shifts to higher temperatures with increasing field as shown with black symbols. Such correlation between the heat capacity and resistivity may indicate that they are both governed by the same physical property related to magnetism.

\begin{figure}[htbp]
\centering
\includegraphics[width=0.47\textwidth]{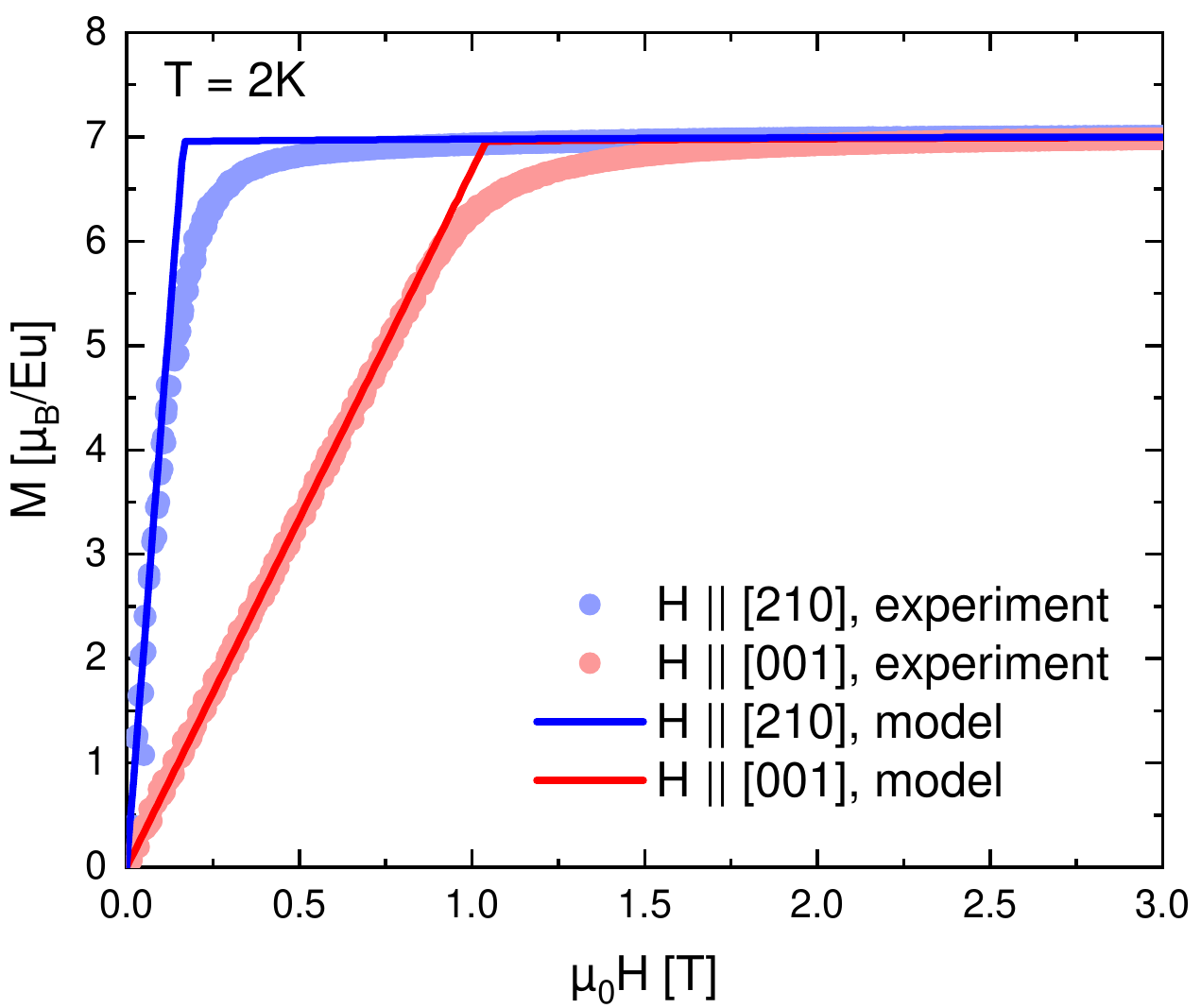}
\caption{\label{fig:EuCd2P2_MvH_2K} Moment versus field at 2\,K for the main symmetry directions in comparison with the model.}
\end{figure}

\subsection{Anisotropic magnetization and static ferromagnetic moment}

The average magnetic moment per Eu$^{2+}$ in dependence of the field, $M(H)$, measured with the field applied along the $[210]$ and $[001]$  directions, Fig.~\ref{fig:EuCd2P2_MvH_2K} (and Fig.~S2 in \cite{SI2023}), reveals the large anisotropy of the material. While for $H\parallel[210]$, saturation is reached already for $B_c^{[210]}=0.17\,\rm T$ at $T=2\,\rm K$, a much higher field, $B_c^{[001]}=1.04\,\rm T$, is needed for field aligned along $[001]$. At $T=2\,\rm K$, for neither of the field directions, a spinflop transition is observed as it can be seen in Fig.~\ref{fig:EuCd2P2_MvH_2K}.  
The difference of the magnetic moment per Eu$^{2+}$, $M_{[210]}-M_{[001]}=M(H\parallel [210])-M(H\parallel[001])$ presents the largest anisotropy at $\mu_0H=0.23\,\rm T$ for $2\,\rm K$ (not shown). Remarkably, while at $2\,\rm K$ the anisotropy almost vanishes above 2\,T, we still observe a significant anisotropy for 10\,K and 20\,K at high fields. The occurrence of the anisotropic shift of the heat capacity maximum for higher fields shown in Fig.~\ref{fig:EuCd2P2_HC_comp}(g) is probably connected to this magnetic anisotropy at higher temperatures.
 
\begin{figure}[ht]
\centering
\includegraphics[width=1\linewidth]{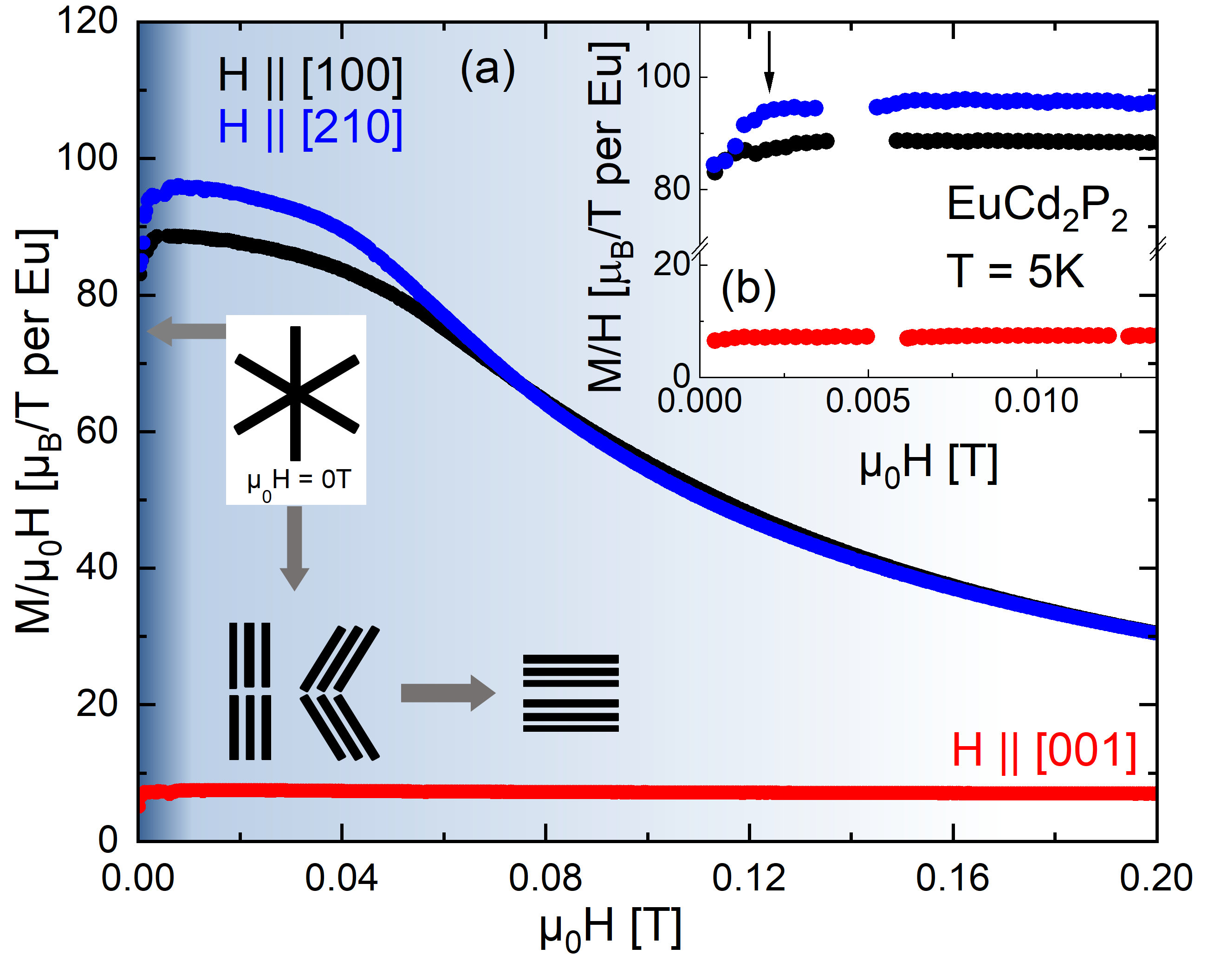}
\caption{(a) $M/H$ vs. $H$ for increasing field at 5 K for the three main symmetry directions. The schematic drawing represents the orientation of the magnetic moments when a field is applied along an in-plane direction. In (b), the low-field region is shown. Reorientation of magnetic domains is observed below $\approx 2.5\,\rm mT$ for in-plane fields (arrows).\label{fig:EuCd2P2_MdurchH}   }
\end{figure}

In Fig.~\ref{fig:EuCd2P2_MdurchH}(a), the data measured at $T=5\,\rm K$ is plotted as $M/H$ versus $H$ at fields below $\mu_0H$ = 0.2\,T. While a strong anisotropy occurs for field along $[001]$ and the two in-plane directions, only a small in-plane anisotropy is visible between the $[100]$ and $[210]$ directions, Fig.~\ref{fig:EuCd2P2_MdurchH}(b). 
EuCd$_2$P$_2$ has an A-type AFM structure with moments in the $a-a$ plane \cite{Wang2021, Sunko2023} which are arranged in six AFM domains in zero field in this trigonal material. By comparing the measured and the simulated data we find that three regimes can be identified for in-plane fields: (i) for fields below $\mu_0H\approx0.0025\,\rm T$ (dark blue shaded area), the moments of the six AFM domains flip perpendicular to the field; (ii) between $0.0025\,\rm T\leq \mu_0H\lesssim 0.1\,\rm T$ (blue shaded area) the moments turn and align along the field; (iii) for $\mu_0H\gtrsim0.1\,\rm T$, the field polarized state is reached.

\begin{figure}[htbp]
\centering
\includegraphics[width=1\linewidth]{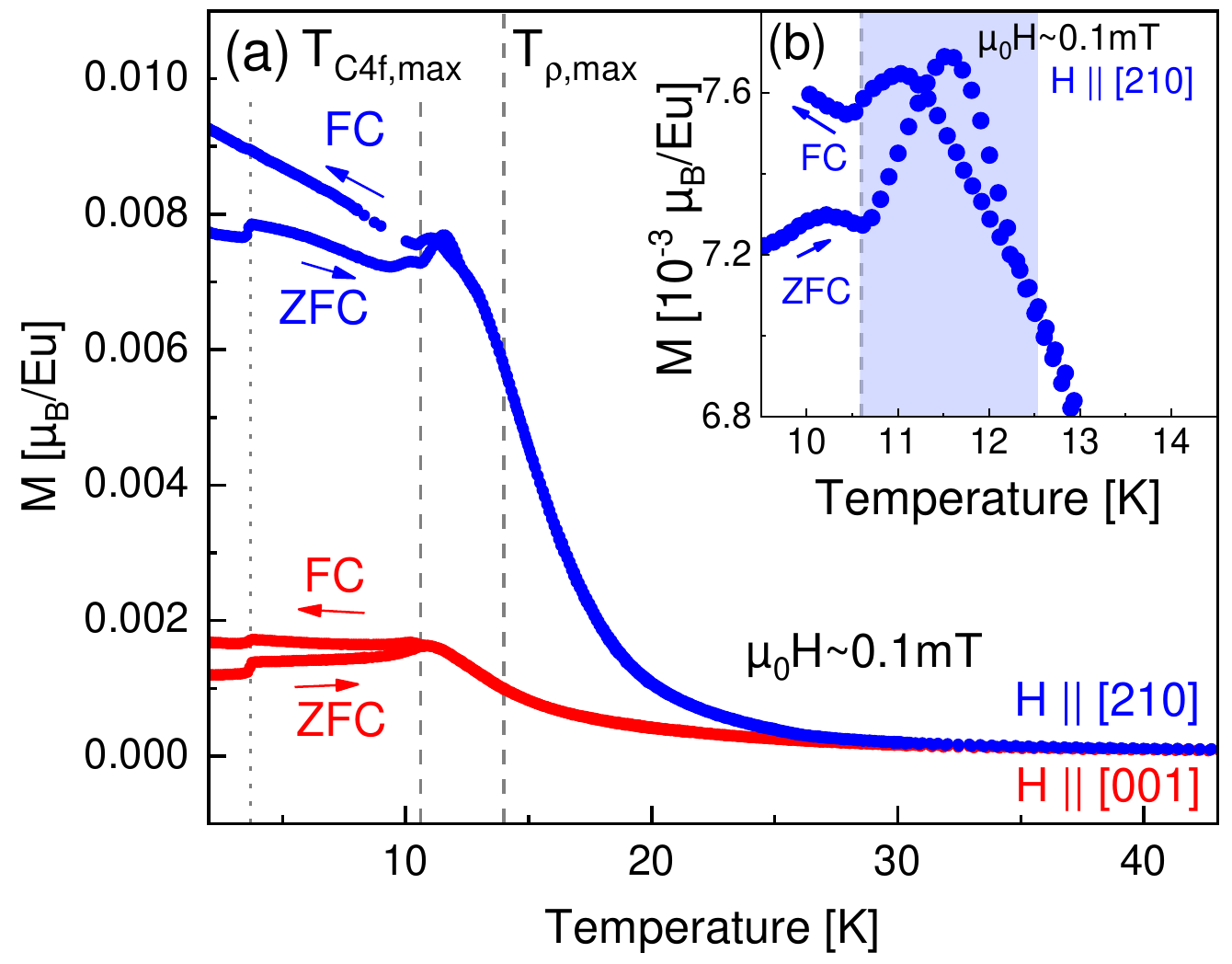}
\caption{\label{fig:inplaneHyst}\label{fig:staticmoment} Magnetic moment versus temperature (a) Zero-field-cooled and field-cooled moment for in-plane and out-of-plane direction at $\mu_0H\approx$ 0.1\,mT\,. Dashed lines mark the temperatures, where the maxima in HC and resistivity appear. The dotted line marks the superconducting transition of tin. (b) Hysteresis effects in the $T$-dependence are observed for both field directions below $\approx 12.5\,\rm K$ (blue shaded area).}
\end{figure}

Considering temperatures still in the AFM regime, slightly below $T_{\rm N}$ we observe nearly no suppression of the resistivity, Fig.~\ref{fig:EuCd2P2_rho_Mueller}, for fields up to $0.3\,\rm T$ which is the field range, where the moments tilt from a perpendicular-to-field to a parallel-to-field orientation, see schematic drawing in Fig.~\ref{fig:EuCd2P2_MdurchH}(b). But we observe a suppression of the resistivity by 1.5 orders of magnitude when increasing the field up to $5\,\rm T$ where the moments are fully polarized in the field.

From the data in Fig.~\ref{fig:EuCd2P2_MdurchH}(b) it is obvious that the energy that is needed to switch the existing AFM domains in the trigonal structure is extremely small as the reorientation of the magnetic domains is observed below very low in-plane fields $\mu_0H\leq 2.5\,\rm mT$ at $T=5\,\rm K$ (black arrow). A similar change of the slope at small fields is also visible at elevated temperatures for $H\parallel[210]$, Fig.~S3) in \cite{SI2023}, meaning that the signatures of magnetic domain reorientation are observed also far above the magnetic ordering temperature up to $T=16\,\rm K$. Interestingly, the feature barely changes when leaving the AFM state. These are the first indications, that magnetically ordered regions are present above $T_N$. For $H\parallel[001]$, we cannot identify a systematic low-field change of slope.

We performed careful measurements of $M(H)$ in the low field region in order to investigate the proposed static ferromagnetic order \cite{Sunko2023}. 
We took remanent fields trapped by the superconducting magnet into account by measuring a Pd standard \cite{PPMS_PdStandard} and found no indications for a ferromagnetic hysteresis in $M(H)$ like the one reported in the supplement of Ref.~\cite{Sunko2023}.
In Fig.~\ref{fig:staticmoment}(a) the static FM moment versus temperature, $M(T)$, is shown.  Close to $T_N$, the moment in presence of a close-to-zero field ($\mu_0H\approx 0.1\,\rm mT$) is approximately four times larger for $H\parallel[210]$ than that for $H\parallel[001]$. Also, for $H\parallel [210]$ the occurrence of a significant contribution even above 20~K is resolved in these data. The inverse magnetic susceptibility, $\chi^{-1}(T)$, measured at $\mu_0H=1\,\rm T$ is shown in Fig.~S4 in \cite{SI2023}. For Eu$^{2+}$ with $L = 0$, no crystalline electric field effects are expected, thus the deviation from the linear fit to the data (fit range $100-290\,\rm K$) according to the Curie-Weiss law is probably connected to the ferromagnetic contribution to the susceptibility. From the data it is obvious that AFM ordering at $\approx 20\,\rm K$ is hindered by the formation of the FM order.
The appearance of a small hysteresis between the ZFC and FC curves in $M(T)$ above $T_N$, below $\approx 12.5\,\rm K$, can be followed up to fields of about 50\,mT, see Figs.~\ref{fig:staticmoment} and ~S5(a),(b) in \cite{SI2023}. Fig.~\ref{fig:staticmoment}(b) shows the enlarged view of the region around the temperature of the peak in the magnetic moment at 0.1~mT. The blue shaded area marks the temperatures between the peak in the heat capacity and that in the resistivity for which the magnetic hysteresis is observed. In Ref.~\cite{Wang2021} the peak in the 0.1~T susceptibility data was taken as a criterion for the determination of the N\'eel temperature. From our data, it is obvious that the peak in the heat capacity at $T_{\rm N}=10.6\,\rm K$ used as the criterion for the AFM ordering temperature does not coincide with the maximum of the magnetic moment but rather with a small dip in the ZFC and FC data in Fig.~\ref{fig:staticmoment}(b).

\begin{figure}[ht]
\centering
\includegraphics[width=0.43\textwidth]{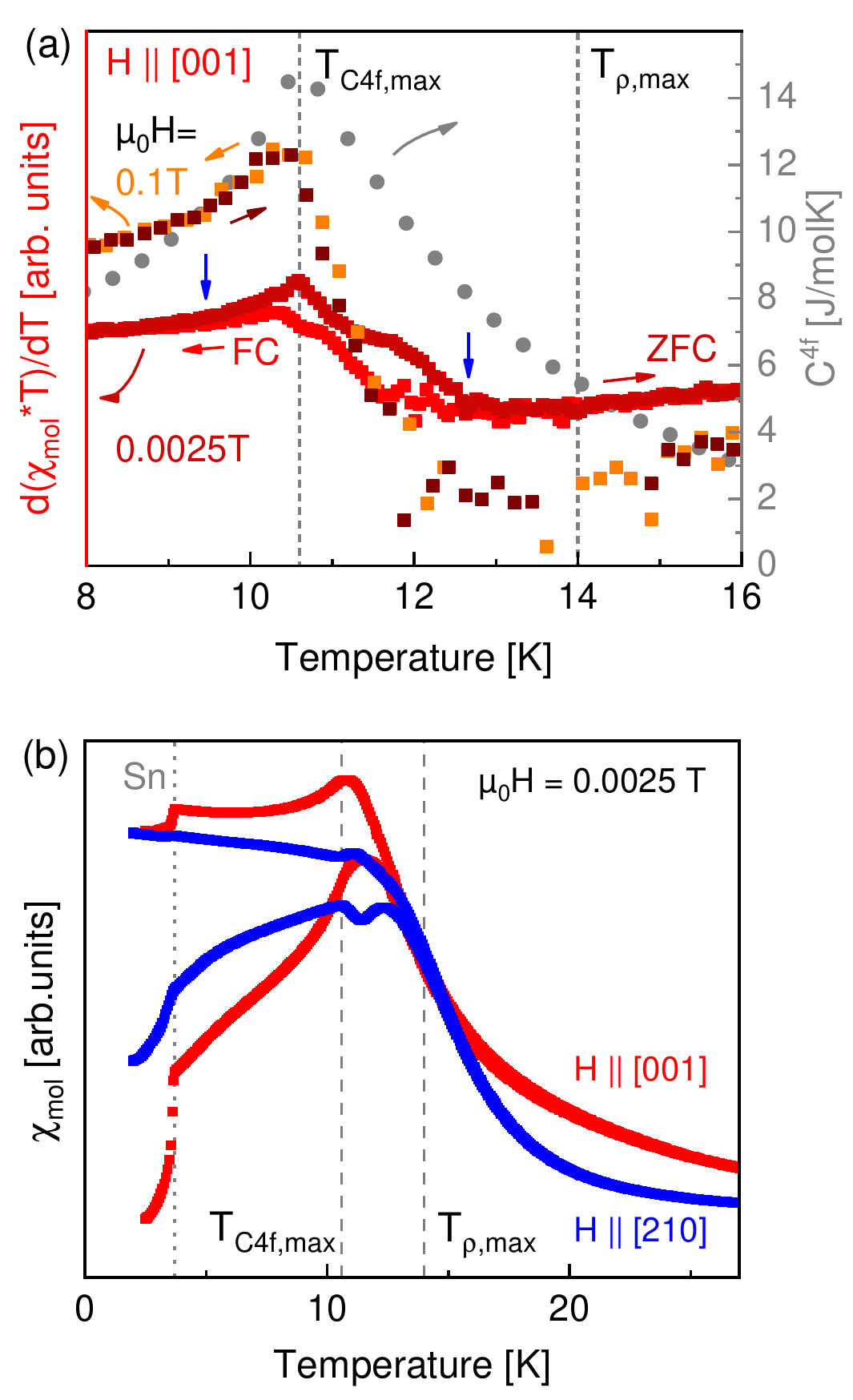}
\caption{\label{fig:EuCd2P2_MvT_Comp_25Oe}\label{fig:EuCd2P2_MvT_Comp_HC} 
EuCd$_2$P$_2$ (a) $d(\chi(T)\cdot T)/dT$ versus temperature measured with $H\parallel [001]$, $\mu_0H=0.0025\,\rm T$ (ZFC, dark red and FC, red) and $\mu_0H=0.1\,\rm T$ (ZFC, brown and FC, orange) in comparison with the magnetic part of the zero field heat capacity, $C^{4f}$. The blue arrows mark the temperatures of the opening and closing of the hysteresis. (b) $\chi(T)$ versus temperature for $H\parallel [001]$ and $H\parallel [210]$ at $\mu_0H=0.0025\,\rm T$ in comparison. The dotted line marks the superconducting transition of Sn. Dashed lines indicate the temperatures where maxima in zero field heat capacity and resistivity appear.}
\end{figure}

\begin{figure*}[htbp]
\centering
\includegraphics[width=0.47\linewidth]{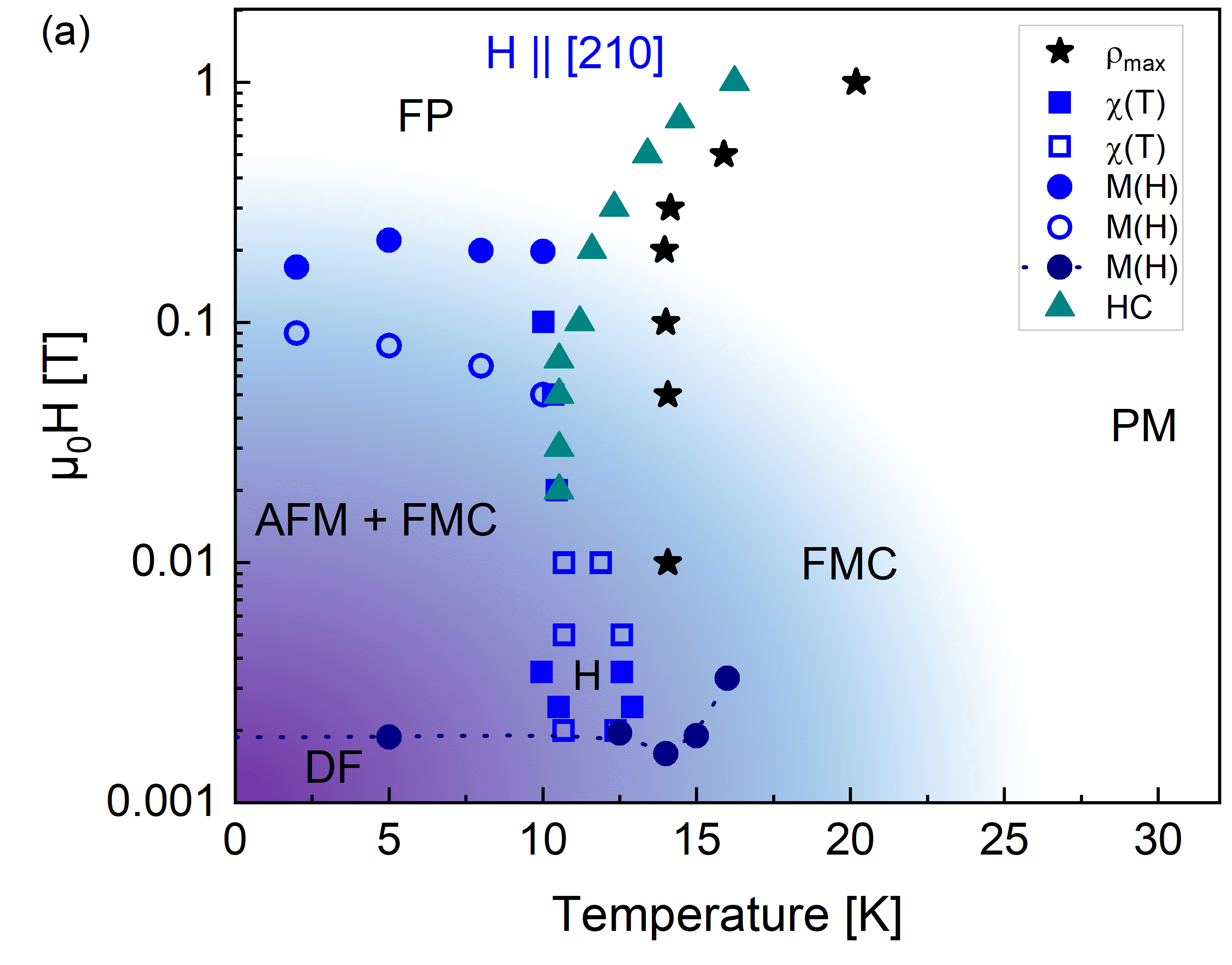}
\includegraphics[width=0.47\linewidth]{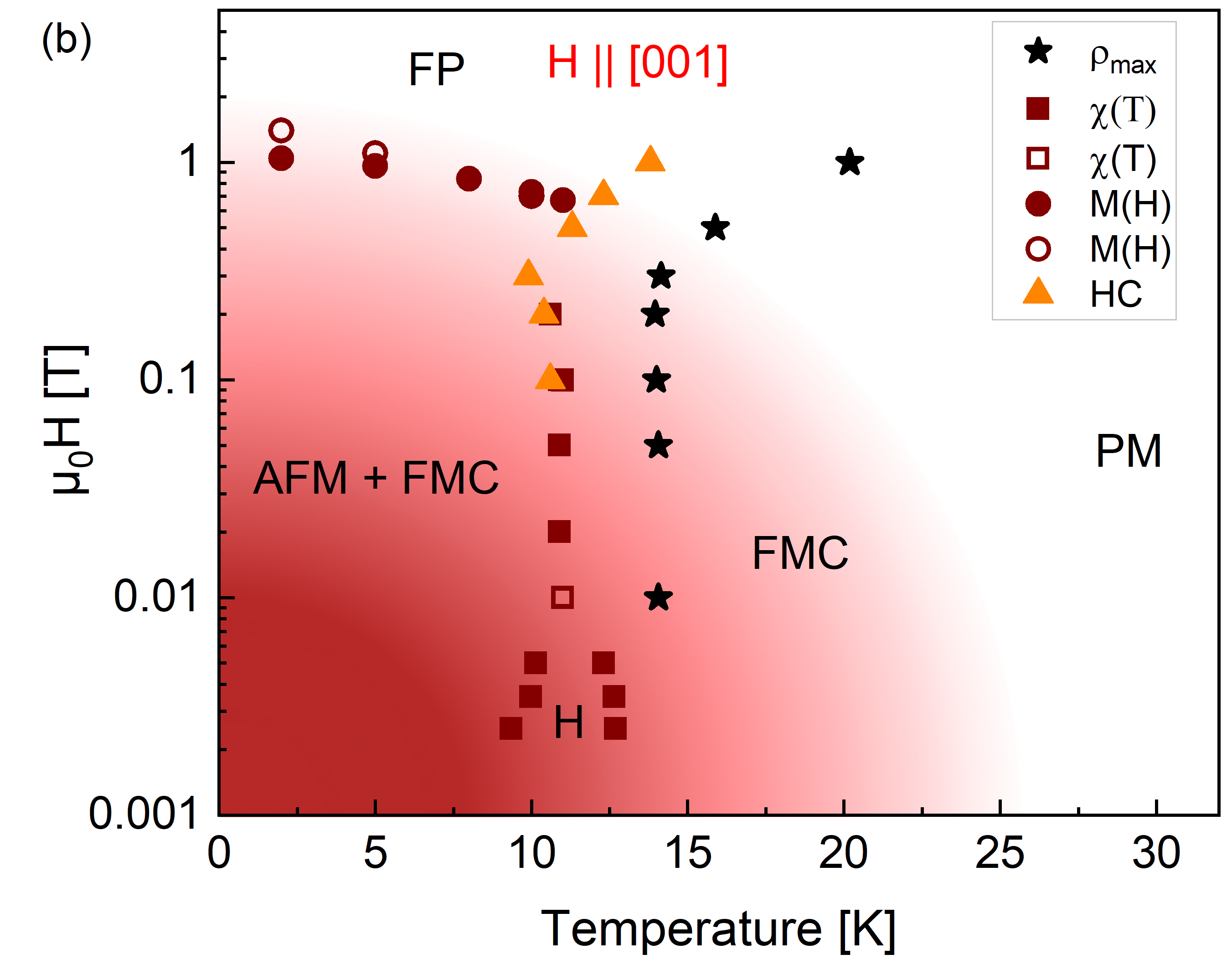}
\caption{\label{fig:PD210} $H-T$ phase diagram of EuCd$_2$P$_2$  for (a) $H\parallel [210]$ and (b) $H\parallel [001]$ constructed from data measured on three different samples. For very low fields and $H\parallel[210]$, we observe signatures of a domain flip (DF) in $M(H)$ (dark blue symbols, data points extracted from Fig.~S3). At low fields, in the H-regime (open and closed square symbols), the opening and closing of the hysteresis of the ZFC and FC susceptibility data were analyzed (blue arrows in  Fig.~\ref{fig:EuCd2P2_MvT_Comp_HC}(a)). For fields higher than $0.01\,\rm T$, closed squares indicate maxima in the susceptibility Figs.~S5, S6 in \cite{SI2023}. Closed and open circles indicate the transition into the field polarized (FP) state in $M(H)$, Fig.~S2 in \cite{SI2023}, which occurs at slightly different fields for different samples. At high temperatures the material is paramagnetic (PM). The blue and red region shows the gradual transition first from PM to the phase where the emergence of a static magnetization is observed which is assigned to the formation of ferromagnetic clusters (FMC). Below T$_N$, the material shows additional  AFM order. Triangles mark the peak temperature in HC and stars the peak temperature in the resistivity for $H\parallel [001]$.} 
\end{figure*}

We performed a more detailed analysis of the magnetic hysteresis in $\chi(T)$ since it is present in the temperature range between the maximum in the heat capacity and that of the resistivity where the occurrence of ferromagnetic clusters was proposed \cite{Sunko2023}. Figure~\ref{fig:EuCd2P2_MvT_Comp_HC}(a) shows a comparison between $d(\chi(T)\cdot T)/dT$ versus temperature (at $\mu_0H=0.0025\,\rm T$ and $\mu_0H=0.1\,\rm T$) and the magnetic part of the heat capacity, $C^{4f}$, in zero field which should correspond to each other from the thermodynamic point of view. $C^{4f}$ was determined from the measured HC subtracting the Debye contribution with Debye temperatures of $\Theta_1$ = 147\,K and $\Theta_2$ = 472\,K (\cite{Anderson2019}, Fig.~S1 in \cite{SI2023}). 
While below $T_{\rm N}=10.6\,\rm K$ the curves show similar characteristics, this is not the case above this temperature. The HC peak shows a significant broadening on its high-temperature flank. 
The above described hysteresis is also visible in $d(\chi(T)\cdot T)/dT$ between the ZFC and FC curves. For $\mu_0H=2.5\,\rm mT$, the hysteresis closes at $T\approx 12.5\,\rm K$, a temperature which is below the peak temperature of the resistivity. The hysteresis, which is visible for very low fields, disappears almost completely for $\mu_0H\geq 0.01\,\rm T$ and is fully closed for $\mu_0H\geq 0.1\,\rm T$ for both field directions. In addition, the characteristics of the susceptibility for the different field directions below $\approx 12.5\,\rm K$ are anisotropic  for $\mu_0H=0.0025\,\rm T$ as it can be seen in Fig.~\ref{fig:EuCd2P2_MvT_Comp_25Oe}(b) which is an indicator for a preferred alignment of the magnetic clusters with respect to the crystal lattice.
We note that this feature occurs in the temperature range where a merging of the putative FM clusters was proposed by \cite{Sunko2023}.
All relevant field and temperature scales in the system are summarized in the $H-T$ phase diagrams in Figs.~\ref{fig:PD210}(a) for $H\parallel [210]$ and (b) for $H\parallel [001]$. The data points were obtained by evaluating the data measured on three different samples and we observe slight sample dependence in the magnetization behaviour when the field is applied along [210] (closed and full blue circles in Fig.~\ref{fig:PD210}(a)). Such small deviations might occur due to different domain distributions, a slight misalignment or even the shape of the sample.
The phase diagrams were constructed from the maxima in resistivity, Fig.~\ref{fig:EuCd2P2_rho_Mueller}, and heat capacity, Fig.~\ref{fig:EuCd2P2_HC_comp}, $M(H)$ and $\chi(T)$ data in Figs.~S2 and S5(a,b) in \cite{SI2023} and from the derivative $d(\chi(T)\cdot T)/dT$ of the data of Figs.~\ref{fig:EuCd2P2_MvT_Comp_HC}(a) and 
Fig.~S3 in \cite{SI2023}. We note that the realignment of the magnetic domains in EuCd$_2$P$_2$ occurs at very low in-plane fields of $\approx 2.5\,\rm mT$ where no significant suppression of the resistivity peak is observed. We find that the domain/magnetic cluster reorientation (phase DF in Fig.~\ref{fig:PD210}(a)) is observed up to temperatures of at least 16~K. In the present case, the change of the domain distribution occurs at very low fields meaning that it requires a very low energy. In other antiferromagnetic systems \cite{Kliemt2017, Kliemt2022, Pakhira2022, Kliemt2023}, similar domain redistribution effects are observed at slightly higher fields of a few hundred Oersteds.

\subsection{Mean-field model for EuCd$_2$P$_2$} 

To simulate the magnetic properties, we used a mean-field linear-chain toy model with two Eu sublattices.  The respective Hamiltonian, which acts only on the $4f$ $|M_J\rangle$ states of Eu, has the form
\begin{multline}
\hat{H}=- \mu_0\sum_{i}^N\left[\vec{\mu}_i\cdot\vec{H}+(1-\gamma)\left(\vec{\mu}_i-\frac{1}{2}\langle\vec{\mu}_i\rangle\right)\vec{M}\right] \\ 
- \frac{3k_B}{J(J+1)} \sum_{i,j}^N T_{ij}\left[\vec{J}_{i}-\frac{1}{2}\langle \vec{J}_{i}\rangle\right]\langle \vec{J}_{j}\rangle - \frac{K_1}{J^2} \sum_{i}^N J_{z,i}\langle J_{z,i}\rangle
\label{eq:H}
\end{multline}
where $N$ is the number of Eu sublattices, $\vec{\mu}_i$ is the magnetic moment operator, $\vec{H}$ is the external magnetic field, $T_{ij}$ are the exchange interaction parameters, $K_1$ defines the single-ion uniaxial anisotropy, $\vec{M}$ is the magnetization, $\gamma$ is the demagnetizing factor, which depends on the shape of the crystal. We used $\gamma=1/3$, as for a spherical object. The second sum is equivalent to the Heisenberg model Hamiltonian, although we used the total angular momenta $J_i$ instead of spins, which differ only in a constant factor. Although the model does not include valence electrons explicitly, the interaction of localized moments with itinerant electrons is taken into account implicitly by including it into the exchange interaction parameters, since exchange interaction between the 4$f$ moments is realized through their coupling to itinerant electrons. Note that our toy Hamiltonian does not include the possibility of FM clusters formation.

The Hamiltonian (\ref{eq:H}) was diagonalized self-consistently using an iterative scheme. We started from several intuitively selected magnetic structures and arrived at a few magnetic configurations from which the one with minimal energy was chosen. The exchange interaction parameters $T_{ij}$ were fitted together with $K_1$ to obtain correct $T_{\rm N}$ and best agreement between the theoretical and experimental dependencies $M(H)$ at 2~K (Fig.~\ref{fig:EuCd2P2_MvH_2K}) for the $[210]$ and $[001]$ directions. The obtained exchange interaction parameters are $T_{11}=T_{22}=10$~K and $T_{12}=T_{21}=-1.23$~K, while the anisotropy parameter is $K_1=-0.35$~meV.

It should be noted that when we switch off the exchange interactions ($T_{ij}=0$), the system exhibits a transition into the FM state at 2~K due to dipole interactions. This temperature is even higher than the AFM exchange coupling between the Eu sublattices ($|T_{12}|=1.23$~K). Thus, dipole FM interactions are in strong competition with the exchange AFM interactions.

\subsection{ARPES and DFT results}

\begin{figure*}[htbp]
\center{\includegraphics[width=\linewidth]{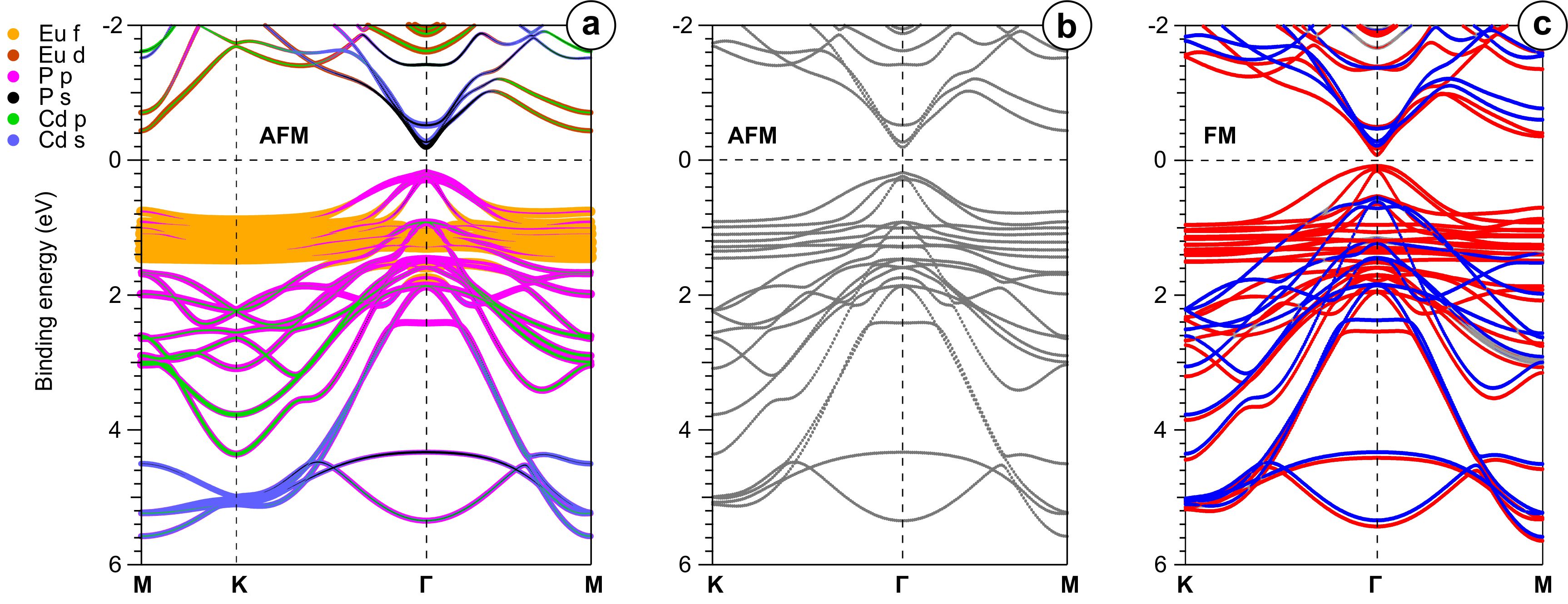}}
\caption{Computed bulk band structure of \ecp\ with (a,b)~A-type AFM and (c)~FM ordering. Colors in (a) highlight the orbital composition of electronic states, while in (b) only the band dispersions are shown. Colors in (c) show the sign of the $S_z$ spin component, while the size of the colored spots denote $|S_z|$; gray spots show the states with small spin polarization.}
\label{fig:1}
\end{figure*}

\begin{figure*}[htbp]
\center{\includegraphics[width=\linewidth]{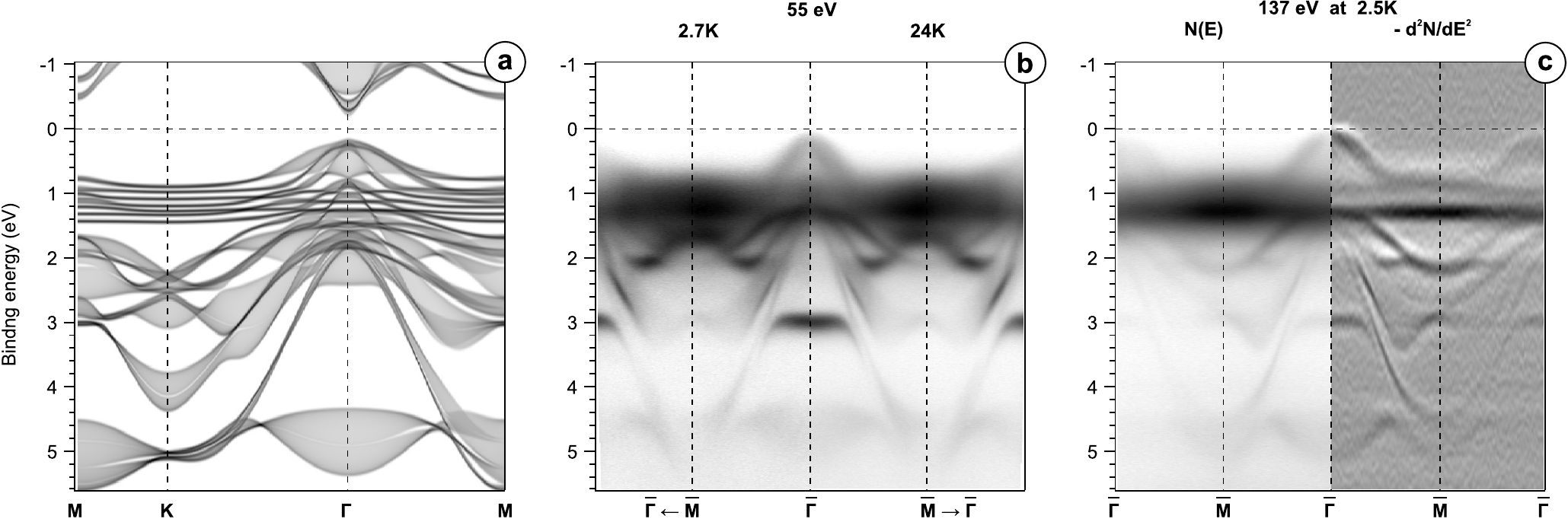}}
\caption{(a)~Calculated AFM bulk electronic states projected on the $(001)$ surface along the $\overline{\text{M}}-\overline{\text{K}}-\overline{{\text{$\Gamma$}}}-\overline{\text{M}}$ direction. Experimental band structure along $\overline{\text{M}}-\overline{{\text{$\Gamma$}}}-\overline{\text{M}}$ direction measured with ARPES using (b)~55~eV and (c)~137~eV photons. }
\label{fig:arpes}
\end{figure*}

We continue with the analysis of electronic structure of EuCd$_2$P$_2$ obtained from \textit{ab initio} calculations and ARPES measurements. Figure~\ref{fig:1} shows the bulk band structures calculated using DFT$+U$ approach for different magnetic configurations. Essentially, the Hubbard parameter $U$ was optimized so that the energy position of the Eu~$4f$ states derived theoretically is consistent with that obtained in the ARPES experiment. The results of the latter are summarized in Fig.~\ref{fig:arpes}. The mentioned consistency was achieved for $U=2.4$ eV. As seen, our calculations reveal the presence of Eu~$4f$ states as narrow bands tightly packed between 0.8 and 1.5~eV binding energy (BE) with almost no dispersion. Because they are located away from the Fermi level, and due to the lifetime of a photo-hole, the individual Eu 4$f$ states cannot be so well resolved in ARPES measurements in comparison, for instance, with those which we resolved for the metallic AFM material EuRh$_2$Si$_2$ \cite{ Hoepner_NC_2013}. 
Our analysis of the orbital composition (Fig.~\ref{fig:1}a) shows that around the $\Gamma$ point the flat Eu 4$f$ states intersect the widely dispersing valence states stemming from Cd and P atoms. Their mutual interaction leads to the emergence of the hole-like $f-p$ hybrid bands that form the apex of the valence band. Note that the similar $f-p$ hybrids have been observed for the AFM semiconductor EuCd$_2$As$_2$~\cite{Ma2019,Soh_PhysRevB_2019,Ma_AdvMater_2020} as well as for the  ferromagnetic semiconductor EuS~\cite{Fedorov_JPCL_2021}, where the hybridization strength and degree of $4f$ localization were found to be crucial for the magnetic exchange interaction and the size of the band gap.

The hybrid states rapidly lose the weight of 4$f$ character on their way to the valence band maximum and finally the latter is formed mainly by the P $3p$ states. On the other hand, the bottom of the conduction band is dominated by the $s$ orbitals of Cd and P. Our calculations indicate that EuCd$_2$P$_2$ is a semiconductor with a direct and narrow band gap at the $\Gamma$ point regardless of the magnetic ordering. In the case of A-type AFM order (Fig.~\ref{fig:1}b), the width of the band gap is about $0.38$~eV. In the case of the FM order the electronic states exhibit a spin splitting due to exchange interaction (Fig.~\ref{fig:1}c). The largest splitting was found near the top of the valence band where it reaches a value of almost 0.5~eV. As a result, the band gap shrinks to $\sim0.16$~eV. Such significant reduction of the gap width may contribute to the CMR effect since the external magnetic field affects the ordering of the moments and may induce a field-polarized FM state with lower resistivity. 
In Fig.~\ref{fig:1} we show the calculations for the case when Eu 4$f$ magnetic moments are oriented along the $c$ axis, since in this case magnetism does not break the hexagonal symmetry of the band structure. More complex case of the in-plane oriented moments is shown in supplemental Fig.~S7.
However, we found that only minor changes in the band structure in the region of the 4$f$ states occur when the moments are oriented in the $a-a$ plane when compared to the out-of-plane orientation. Thus, the anisotropy of the transport and thermodynamic properties is not related to the changes of the band structure upon rotation of the magnetic field direction. More likely, it is related to the anisotropy of magnetic interactions and spin fluctuations. For the FM calculation, we used the same hexagonal unit cell as in the AFM calculation (not the primitive one) for better evaluation and comparison of the bands.

It is worth noting that the electronic structure of {\ecp} is very similar to that of {\ezp} studied recently ~\cite{Krebber2023}. However, there is one essential difference. DFT calculation predicts a direct gap for {\ecp}, while for {\ezp} it was shown both theoretically and experimentally that the gap is indirect. Our calculation shows that the conduction band minimum for {\ezp} is located at the $M$ point of the Brillouin zone. To explain such a difference between the two materials, we performed DFT calculations of {\ecp} with the lattice parameters of {\ezp}. For such significantly strained material, we obtained the gap in the $\Gamma$ point of 0.46~eV, while the indirect gap of 0.22~eV was predicted between the valence band states in $\Gamma$ and conduction band states in the $M$ point. The latter value is very similar to the predicted indirect gap width of 0.2~eV in {\ezp}. These results disclose the potential for application of strain for handling the fundamental band gap in semiconducting materials and namely, for manipulation of the direct and indirect gap and its value in this material class.

Let us now focus at the ARPES study of the band structure. Figure~\ref{fig:arpes}b presents the ARPES data acquired from the cleaved $(001)$ surface of the {\ecp} crystal at two temperatures above and below $T_{\rm N}$. The measurements were performed at various temperatures in the range 2.5--30~K and for different excitation energies. Note that we did not detect any significant changes in the spectral pattern of the electron bands. In Figure~\ref{fig:arpes}b we show the ARPES data taken at 2.5~K and 24~K, while in the supplemental Fig.~S8, we present also the results obtained at 14~K. 

The experimental spectra can be best compared to the calculated AFM bulk band structure projected along the $k_z$ direction that is shown in Fig.~\ref{fig:arpes}a. At the $\Gamma$ point we can see several hole-like bands originated mainly from the $p$-orbitals of phosphorus according to Fig.~\ref{fig:1}a. The overall experimental band structure shows good agreement with the calculated one, however the flat band near the $\Gamma$ point that appears at 3~eV in the experiment is shifted by about $0.5$~eV in the calculation. The bands with higher binding energy (above $3$~eV) are more pronounced in the ARPES spectra acquired with the photon energy of $137$~eV, which is shown in Fig.~\ref{fig:arpes}c together with its second derivative.

In agreement with the DFT$+U$ calculations, the ARPES measurements clearly show semiconducting properties with the Fermi level located above the valence band maximum. We have to emphasize that no traces of the step related to the Fermi function were detected at any temperature. Figure~\ref{fig:arpes}b shows that the band dispersions below and above the magnetic transition temperature remain almost identical. No exchange splitting of bands is observed at any temperature pointing to the AFM ordering of moments in the magnetically ordered phase. Thus our results are in contradiction with the recently appeared ARPES data~\cite{zhang2023electronic} that demonstrate metallic properties of {\ecp} at low temperatures by revealing the bands crossing the Fermi level. Also, Ref.~\cite{zhang2023electronic} reports on significant changes of the band structure near the top of the valence bands upon changing the temperature from 26 to 6~K. The results of our synchrotron-based ARPES measurements do not support these results, and the origin of this discrepancy remains unclear.

\section{Conclusions}

In summary, we have grown and comprehensively characterized high-quality single-crystals of EuCd$_2$P$_2$ with the size of up to $3\times3\times1$~mm$^3$. The transport measurements ultimately indicate a strong peak in resistivity at 14~K and not at 18~K as it was reported recently. Further, we focused on the heat capacity measurements and characterization of the magnetic properties of EuCd$_2$P$_2$. The Curie-Weiss analysis of the inverse susceptibility shows that the system is expected to enter the magnetically ordered phase at $\sim20$~K, which is essentially above its $T_{\rm N}$ of $10.6$~K. Thus, in the temperature range of $10.6-20$~K the magnetic ordering is suppressed and at that the resistivity peak is observed.
The measurements of resistivity and heat capacity as a function of applied magnetic field revealed correlated changes in the positions of the characteristic peaks in these dependencies. This indicates that the resistivity is connected to the behavior of the Eu $4f$ magnetic moments, that dominates the heat capacity. For the fields above 0.3~T we observed a shift of the heat capacity peak towards higher temperatures. The shift is significantly larger than that predicted by the mean-field model for an AFM material. This points to the increase of ferromagnetic interactions above $T_{\rm N}$. Our mean-field model also indicates that AFM coupling in this system is even weaker than the dipole FM interaction implying that the two interactions are in strong competition. Performing ZFC and FC $\chi(T)$ measurements, we detected a hysteresis above $T_{\rm N}$ up to $\approx 12.5$~K. It indicates that some FM ordering is present at the low-T side of the resistivity peak. We furthermore observed a reorientation of magnetic domains at temperatures up to $\approx 2\,T_{\rm N}$ confirming the presence of magnetic order.
This is in qualitative agreement with the results of Ref.~\cite{Sunko2023} indicating the presence of static magnetization above $T_{\rm N}$. 
More importantly, we find that the static moment is four times larger in the in-plane compared to the out-of-plane direction indicating a magnetocrystalline anisotropy which may also lead to a preferred alignment of the magnetic clusters above $T_{\rm N}$. 
Our $\chi(T)$ data obtained at $2.5\,\rm mT$ demonstrate that the ZFC-FC hysteresis occurs within the temperature range between heat capacity maximum and the temperature of the peak in the resistivity. It was proposed in \cite{Sunko2023}, that in this temperature range the percolation of the FM clusters appears. 
In contrast to the results of Ref.~\cite{Sunko2023} (supplement), we did not observe a ferromagnetic hysteresis in  $M(H)$.

We can rule out the reorientation of magnetic domains being the main mechanism of the CMR effect, but observe that the strongest suppression of the resistivity peak is observed for fully field-polarized moments.
To characterize the electronic structure of EuCd$_2$P$_2$ and to shed some light on possible metallicity reported in Ref.~\cite{zhang2023electronic} at low temperatures, we performed synchrotron-based ARPES measurements in the temperature range of $2.5-24$~K and DFT$+U$ band structure calculations. Essentially, we did not reveal any notable changes of the band structure with temperature and we did not detect any indications of metallicity of EuCd$_2$P$_2$ down to 2.5~K. Our ARPES data confirm that EuCd$_2$P$_2$ is an AFM semiconductor, however, they do not exclude possible local FM order that does not significantly change the total density of states.

\section{Acknowledgments}
We acknowledge funding by the Deutsche Forschungsgemeinschaft (DFG, German Research Foundation) via the SFB/TRR 288 (422213477, project A03, B02) and via the SFB1143 (47310070, project C04).
We acknowledge the support from the Saint Petersburg State University (Grant No. 95442847). 
The work of D.Yu.U. was partially supported by the Ministry of Science and Higher Education of the Russian Federation (No. FSMG-2023-0014) (modeling of magnetism) and RSF 23-72-30004 (electronic structure analysis).
We acknowledge the Helmholtz-Zentrum Berlin f\"ur Materialien und Energie for beamtime allocation at the One-cube ARPES instrument at the BESSY~II synchrotron.

\end{document}